\documentclass[preprint]{aastex62}
\usepackage{graphicx}
\usepackage{url}                                                              
\usepackage[]{natbib}     
\usepackage{epsfig}
\usepackage{color}
\usepackage{xspace}

\definecolor{DarkGreen}{rgb}{0.0,0.4,0.0}  

\submitjournal{ApJ}
\shorttitle{Double-Decker Filament Configuration}
\shortauthors{Awasthi, Liu \& Wang}

\begin{document}

\title{Double-Decker Filament Configuration Revealed by Mass Motions}

\author[0000-0001-5313-1125]{Arun Kumar Awasthi}
\affiliation{CAS Key Laboratory of Geospace Environment, Department of Geophysics and Planetary Sciences, University of Science and Technology of China, Hefei 230026, China}
\affiliation{CAS Center for Excellence in Comparative Planetology, Hefei 230026, China}
\author[0000-0003-4618-4979]{Rui Liu}
\affiliation{CAS Key Laboratory of Geospace Environment, Department of Geophysics and Planetary Sciences, University of Science and Technology of China, Hefei 230026, China}
\affiliation{CAS Center for Excellence in Comparative Planetology, Hefei 230026, China}
\author[0000-0002-8887-3919]{Yuming Wang}
\affiliation{CAS Key Laboratory of Geospace Environment, Department of Geophysics and Planetary Sciences, University of Science and Technology of China, Hefei 230026, China}
\affiliation{CAS Center for Excellence in Comparative Planetology, Hefei 230026, China}
\correspondingauthor{Rui Liu}
\email{rliu@ustc.edu.cn}

\begin{abstract}
It is often envisaged that dense filament material lies in the dips of magnetic field lines belonging to either a sheared arcade or a magnetic flux rope. But it is also debated which configuration correctly depicts filaments' magnetic structure, due to our incapacity to measure the coronal magnetic field. In this paper, we address this issue by employing mass motions in an active-region filament to diagnose its magnetic structure. The disturbance in the filament was driven by a surge initiated at the filament's eastern end in the NOAA active region 12685, which was observed by the 1-m New Vacuum Solar Telescope (NVST) in the H$\alpha$ line center and line wing ($\pm0.4$~{\AA}). Filament material predominately exhibits two kinds of motions, namely, rotation about the spine and longitudinal oscillation along the spine. The former is evidenced by antisymmetric Doppler shifts about the spine; the latter features a dynamic barb with mass extending away from the H$\alpha$ spine until the transversal edge of the EUV filament channel. The longitudinal oscillation in the eastern section of the filament is distinct from that in the west, implying that the underlying field lines have different lengths and curvature radii. The composite motions of filament material suggest a double-decker host structure with mixed signs of helicity, comprising a flux rope atop a sheared-arcade system.
\end{abstract}


\section{Introduction}
Solar filaments are composed of dense (10$^{11-12}$ cm$^{-3}$) and cold plasma (10$^4$K) suspended in the tenuous (10$^{8-9}$ cm$^{-3}$) and hot (10$^6$K) corona, hence appear dark in H$\alpha$ or EUV against the solar disk. As filament threads are most likely aligned with magnetic field \citep{Lin2005}, high-resolution observations possess the potential of disclosing magnetic structures above the photosphere, which we still cannot measure directly. In EUV, a dark corridor termed ``EUV filament channel'' is well extended beyond the H$\alpha$ filament. This enhanced width is explained by Lyman continuum absorption of EUV radiation ($\lambda< 912$~{\AA}) and ``volume blocking'', an additional reduction in EUV intensity because the cool plasma occupying the corridor does not emit any EUV radiation \citep{Anzer+Heinzel2005}. Usually solar eruptions originate from filament channels, where dense filament material may or may not be present but the magnetic field is highly non-potential and opposite in polarity at two sides. Hence the magnetic configuration of filaments is crucial for understanding the physics of solar eruptions \citep{Mackay2010}.

Three kinds of filament models with distinct magnetic configurations have been proposed in the literature, namely the wire model \citep{Martin1994, Lin2008}, the sheared arcade model \citep{Kippenhahn1957}, and the twisted flux-rope model \citep{Kuperus1974}. The empirical wire model assumes that a filament is composed of field-aligned fine threads. It differs from the other two in the absence of magnetic dips, which are present either at the top of a sheared arcade or the bottom of a flux rope. Besides its structural and morphological similarities with coronal cavities \citep[\S3.3 in][]{Gibson2018}, the flux rope model is appealing in that its helical windings provide for filament plasma both support against gravity and thermal insulation from the hot corona. Further, it explains the inverse-polarity configuration observed often in quiescent filaments \citep{Leroy1984, Bommier1998}, i.e., the magnetic field traversing the filament is directed from negative to positive polarity. The sheared arcade model generally implies a normal-polarity configuration, but a mixture of normal- and inverse-polarity dips is found in numerical experiments \citep{Aulanier2002}.

Although the flux-rope model is consistent with many active-region filaments   \citep[e.g.,][]{Dudik2008,Canou2010,Sasso2014,LiuR2014,Liu2016}, more complicated magnetic configurations are not rare. For example, \citet{Guo2010} found that a flux rope and a sheared arcade match two sections of a filament separately. To explain a `double-decker' filament that was resolved stereoscopically and later erupted partially, \citet{Liu2012} proposed two possible configurations, either a double flux rope or a single flux rope atop a sheared arcade. The former configuration has found support in modeling \citep{Kliem2014} and in more observations \citep[e.g.,][]{Cheng2014,Zhu2015,Liu2016,Liu2017}. The latter configuration is considered natural for a flux rope suspended high in the corona, but direct observational evidence has been lacking. Further, \citet{Awasthi2018} identified an even more complex flux-rope system with multiple branches braiding about each other, displaying signatures of internal reconnections.

An additional modeling constraint is the pattern of filament chirality, which is consistent with the hemispheric rule of magnetic helicity \citep{Pevtsov1995}. By definition, a filament is \emph{dextral} (\emph{sinistral}) if its axial magnetic field points right (left) when viewed from its positive-polarity side. It is believed that a dextral (sinistral) filament has right-bearing (left-bearing) barbs, a bundle of filament threads extruding out of the filament spine in a way similar to right- or left-bearing exit ramps off a highway. The majority of filaments in the northern (southern) hemisphere indeed have right-bearing (left-bearing) barbs and are overarched by left-skewed (right-skewed) coronal arcades, corresponding to the dominantly negative (positive) helicity in the same hemisphere \citep{Martin1998, Bernasconi2005, Pevtsov2003, Yeates2007}. However, \citet{Chen2014} argued that the correspondence between the filament chirality and the bearing sense of barbs works only for filaments supported by flux ropes and the correspondence is reversed for sheared arcades, if the sheared arcade possesses the same sign of helicity as the flux rope. Alternatively, \citet{Chen2014} proposed that a filament is dextral (sinistral) if during the eruption the conjugate sites of plasma draining are right-skewed (left-skewed) with respect to the polarity inversion line (PIL). With this new criteria, the hemispheric rule of filament chirality is significantly strengthened \citep{Ouyang2017}.

Mass motions in a filament also provide clues on its host magnetic field \citep[e.g.,][]{Zirker1998,Okamoto2016,Wang2018} or how it interacts with the surrounding field \citep[e.g.,][]{Liu2018}, assuming a low-$\beta$ plasma environment. Filament `winking' \citep{Dyson1930}, later recognized as large-amplitude oscillations \citep{Ramsey1966}, is promising in probing the filament magnetic field. Often activated by shock waves impinging on the filament side, transverse oscillations perpendicular to the filament spine have been modeled by a damped harmonic oscillator with magnetic tension serving as the restoring force \citep{Hyder1966}. In contrast, longitudinal oscillations along the spine are often activated by a subflare \citep[e.g.,][]{Jing2003,Jing2006} or a jet \citep[e.g.,][]{Luna2014} at one end of the filament, or, occasionally by a shock wave propagating along the filament spine \citep[e.g.,][]{Shen2014}. Various restoring forces have been considered since the discovery of the phenomenon \citep{Jing2003}, e.g., the magnetic pressure gradient \citep{Vrvsnak2007}, the gas pressure gradient \citep{Jing2003,Vrvsnak2007}, and the projected gravity in a magnetic dip \citep{Jing2003,ZhangQM2012,Luna+Karpen2012}. The first two forces have implications that are seldom observed, either predicting motions perpendicular to the local magnetic field \citep[however, see][]{Zhang2017transverse} or requiring a temperature difference of several million Kelvins \citep{Vrvsnak2007}. The simplified pendulum model, however, appears self-consistent and can provide diagnostics on magnetic parameters such as the curvature of the field-line dip and the minimum field strength \citep{Luna+Karpen2012,Luna2014}.

Here we investigate in an active-region filament two co-existing dynamic motions triggered by a small flare, i.e., longitudinal oscillations along, and rotary motions about, the spine (\S\ref{sec:obs}). In \S\ref{sec:disc_conc} we confront the existing filament models with the observations and propose a double-decker configuration.

\section{Observations \& Analysis} \label{sec:obs}
\subsection{Instruments}
In this study we mainly used H$\alpha$ observations acquired by the 1-m New Vacuum Solar Telescope (NVST; \citet{LiuZ2014}) at three wavelength positions, namely the line-center and the line wing at $\pm$0.4~{\AA}, during 02:10:54 UT--04:57:02 UT on 2017 October 26. After being processed by a speckle-masking method, the H$\alpha$ images have a pixel scale of $0''.136$ and a cadence of 28 s. To provide the context, we have also utilized EUV images of the full-disk Sun with a pixel scale of $0''.6$ (corresponding to a spatial resolution of $1''.5$) and a temporal cadence of 12~s acquired by the Atmospheric Imaging Assembly \citep[AIA;][]{Lemen2012}) onboard \textit{Solar Dynamical Observatory} \citep[SDO;][]{Pesnell2012}, photospheric magnetograms made available by the Helioseismic and Magnetic Imager \citep[HMI;][]{Scherrer2012}) onboard \textit{SDO}, full-disk H$\alpha$ images with an $1''$ pixel scale and a 60-s temporal cadence provided by the \emph{Global Oscillation Network Group (GONG)}, and soft X-ray flux at 1--8 and 0.5--4~{\AA} by the \textit{Geostationary Operational Environmental Satellites (GOES)}.

\subsection{Morphological and Dynamical Characteristics of the Filament}\label{sec:results}
\begin{figure}
\centering
\includegraphics[height=0.99\textwidth]{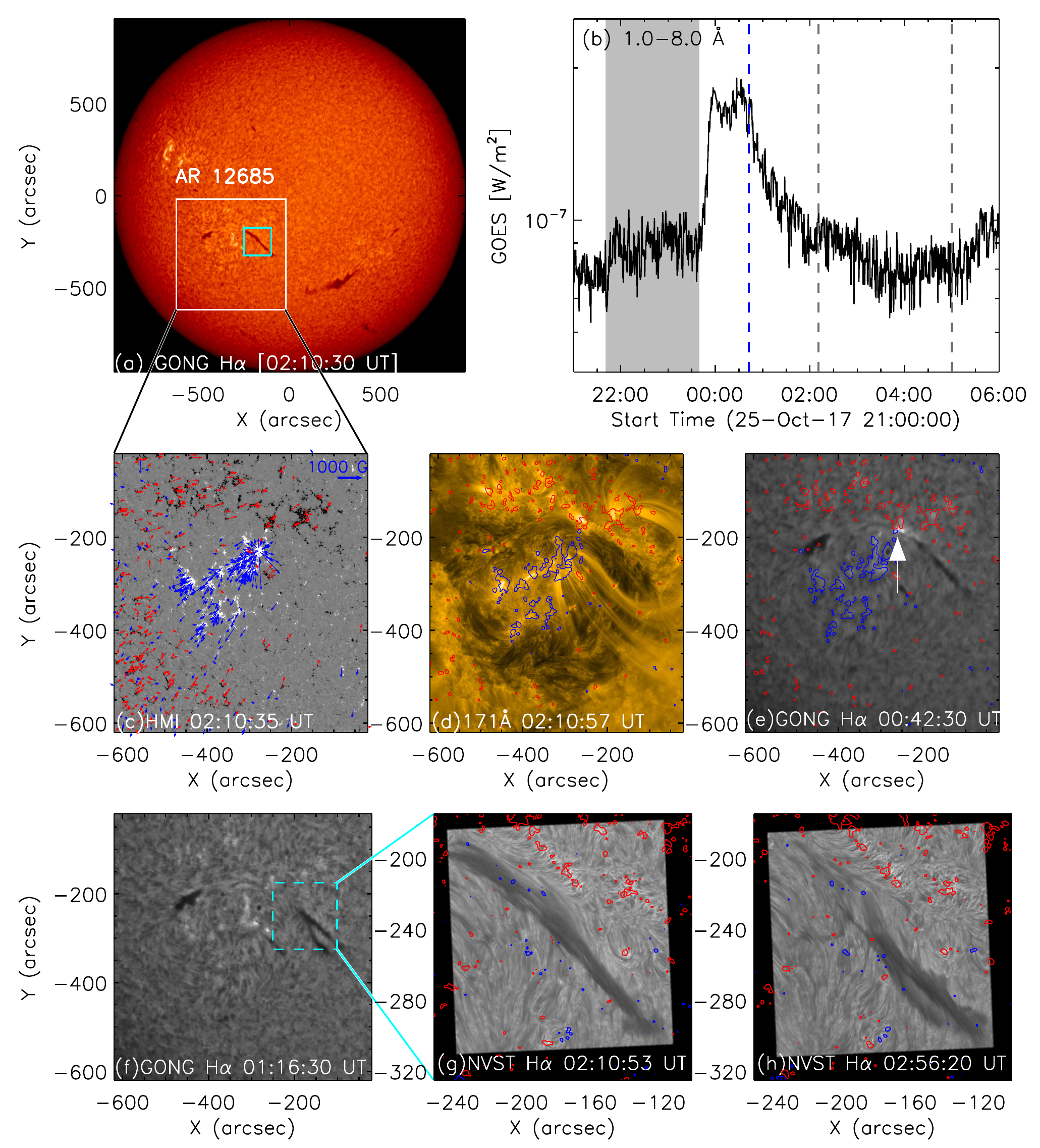}
\caption{Multi-wavelength overview of the filament. (a) \emph{GONG} full-disk Sun in H$\alpha$ denoting the limited NVST coverage (cyan rectangle) of the full active region (white rectangle). (b) \emph{GOES} 1--8 and 0.5--4~{\AA} light curve showing a B1.9-class flare. The blue dashed line marks the time of the surge, and the grey dashed lines mark the NVST observing duration during the decay phase of the flare. The shaded region denotes the co-spatial precursor activities. A dark EUV filament channel of circular shape is located along the PIL (c \& d). Its northern section is occupied by filament segments in H$\alpha$ (e). Following the flare, a dynamic barb formed by oscillatory filament material in H$\alpha$ (f--h). The line-of-sight (LOS) magnetogram (c) is saturated at $\pm100$~G and superimposed by the transverse component exceeding 200~G. The blue (red) contours superposed upon the AIA 171~{\AA} image (d) and H$\alpha$ images (e, g \& h) denote positive (negative) LOS component of the photospheric magnetic field at 50~G. In each frame of the accompanying online movie\textsuperscript{a}, the top panels show the GOES light curve and LOS HMI magnetogram, and the bottom panels show an AIA 171~{\AA} image, a GONG H$\alpha$ and an NVST H$\alpha$ image.} \label{fig:filament_and_flare}
\small\textsuperscript{a} Available at \url{https://www.dropbox.com/s/0bcskjgy5fmn5sp/aia_gong_nvst_movie.avi?dl=0}
\end{figure}

The filament of interest was located in the NOAA active region 12685 on 2017 October 26 (Figure~\ref{fig:filament_and_flare}a). In the line-of-sight magnetogram, the active region is characterized by positive magnetic fluxes around the sunspot in the center and negative fluxes in the surrounding facular region (Figure~\ref{fig:filament_and_flare}c). Aligned along the PIL, the dark EUV filament channel possesses a similar circular shape, as observed in 171~{\AA} (Figure~\ref{fig:filament_and_flare}d). The H$\alpha$ filament under investigation occupies the northwestern section of the filament channel, oriented in a northeast-southwest direction (Figure~\ref{fig:filament_and_flare}e). For simplicity, the filament's two half sections are hereafter referred to as the eastern and western section, respectively.

The filament was already dynamic in the earliest image recorded by NVST (Figure~\ref{fig:filament_and_flare}g), which covered the gradual phase of a B1.9-class flare (Figure~\ref{fig:filament_and_flare}b). Analyzing earlier AIA 171~{\AA} and \emph{GONG} H$\alpha$ observations, we found that particularly the eastern section of the H$\alpha$ filament under investigation exhibits multiple episodes of disturbances energized by precursor activities near the eastern end of the filament, where positive fluxes in the south contact negative fluxes in the north (marked by an arrow in Figure~\ref{fig:filament_and_flare}c). The B-class flare occurs in the same place and initiates a surge (Figure~\ref{fig:filament_and_flare}e and accompanying animation) that is directed westward and disturbs the filament. Following the flare, filament material in the eastern section of the H$\alpha$ filament shoots southwestward, reaching as far as $\sim\,$20$''$ away from the spine, and then moves backward, forming a `dynamic' barb structure that is right-bearing with respect to the filament spine (Figure~\ref{fig:filament_and_flare}(g--h)). Below the dynamical motions are investigated in detail.

\subsubsection{Large-Amplitude Longitudinal Oscillations}
\begin{figure}
\centering
\includegraphics[width=0.99\textwidth, angle=0]{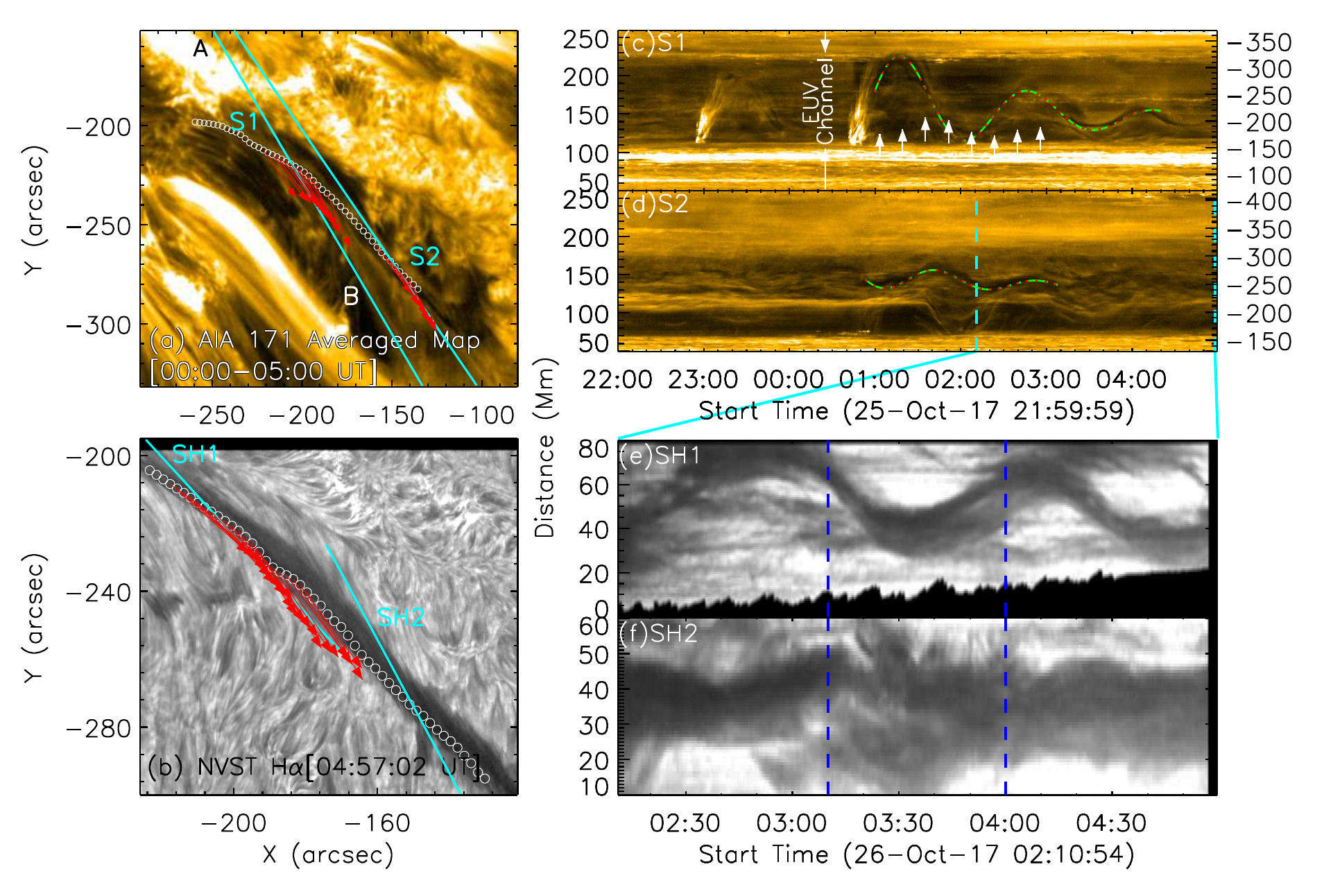}
\caption{Large-amplitude longitudinal oscillation of the filament material. The filament spine is represented by 50 uniformly-spaced points (white circles) identified visually from the averaged 171 $\mathrm{\AA}$ image during 00:00--05:00 UT in Panel (a) and the H$\alpha$ image at 04:57:02 UT in Panel (b). Red arrows in (a) and (b) point towards the determined direction of oscillating material (see text). The TD maps corresponding to the two slits S1 and S2 (marked in (a), in the direction A to B) are plotted in (c) and (d), respectively. Similarly, TD maps prepared from the H$\alpha$ images corresponding to the slits SH1 and SH2 (marked in (b)) are plotted in (e) and (f), respectively. The time interval of (e \& f) is marked in (d) by cyan dashed lines. The slits extend beyond the FOV for which the EUV and H$\alpha$ images are shown. The best-fit damped sinusoidal (red) and Bessel (green) curves are overplotted in the EUV TD maps. The oscillation appears to be constrained within the EUV channel as annotated in (c). In the eastern section of the filament, a faint yet discernible low-amplitude oscillation can be seen in the TD map (Panel (c), marked by a sequence of arrows), in addition to the dominant oscillation pattern. Blue vertical lines on (e) and (f) indicate the time of the Doppler maps in Figure~\ref{fig:filament_rotation}(a and b).}
\label{fig:aia_nvst_oscillation}
\end{figure}

To characterize the oscillations of the filament, we employed the technique of \citet{Luna2014, Luna2018}. Firstly, the filament spine is represented by 50 uniformly-spaced points picked by visual inspection of the averaged EUV 171 $\mathrm{\AA}$ image (during 00:00--05:00 UT; Figure~\ref{fig:aia_nvst_oscillation}a) and NVST H$\alpha$ image obtained at 04:57:02 UT (Figure~\ref{fig:aia_nvst_oscillation}b), respectively. We then constructed time-distance maps (hereafter TD maps) through a series of virtual slits centered on each spine point and oriented with angles relative to the solar west ranging from +15$^{\circ}$ to -85$^{\circ}$ at a step of 2$^\circ$. The slits are numbered 1--50 from east to west. A total of 2500 TD maps were prepared from EUV 171 $\mathrm{\AA}$ images obtained during 25-Oct-2017 21:00 UT--26-Oct-2017 05:00 UT (in view of the prolonged precursor activities; see Figure~\ref{fig:filament_and_flare}b) and all the NVST H$\alpha$ images available. The slit length is $360''$ (261 Mm) for EUV (Figure~\ref{fig:aia_nvst_oscillation}(c \& d)) and $108''$ ($\sim$79 Mm) for H$\alpha$ images (Figure~\ref{fig:aia_nvst_oscillation}(e \& f)). The TD maps display typical large-amplitude longitudinal oscillations, with temporal agreement between EUV and H$\alpha$. EUV TD maps reveal the flare-induced brightening as the trigger of the filament oscillation, which reaches as far as the boundary of the EUV filament channel (Figure~\ref{fig:aia_nvst_oscillation}c). Further, a precursor surge at about 23:00 UT on 2017 October 25 also triggered the oscillation in the same direction as that of the impending flare, however with a smaller amplitude.

\begin{figure}
\centering
\includegraphics[width=0.99\textwidth, angle=0]{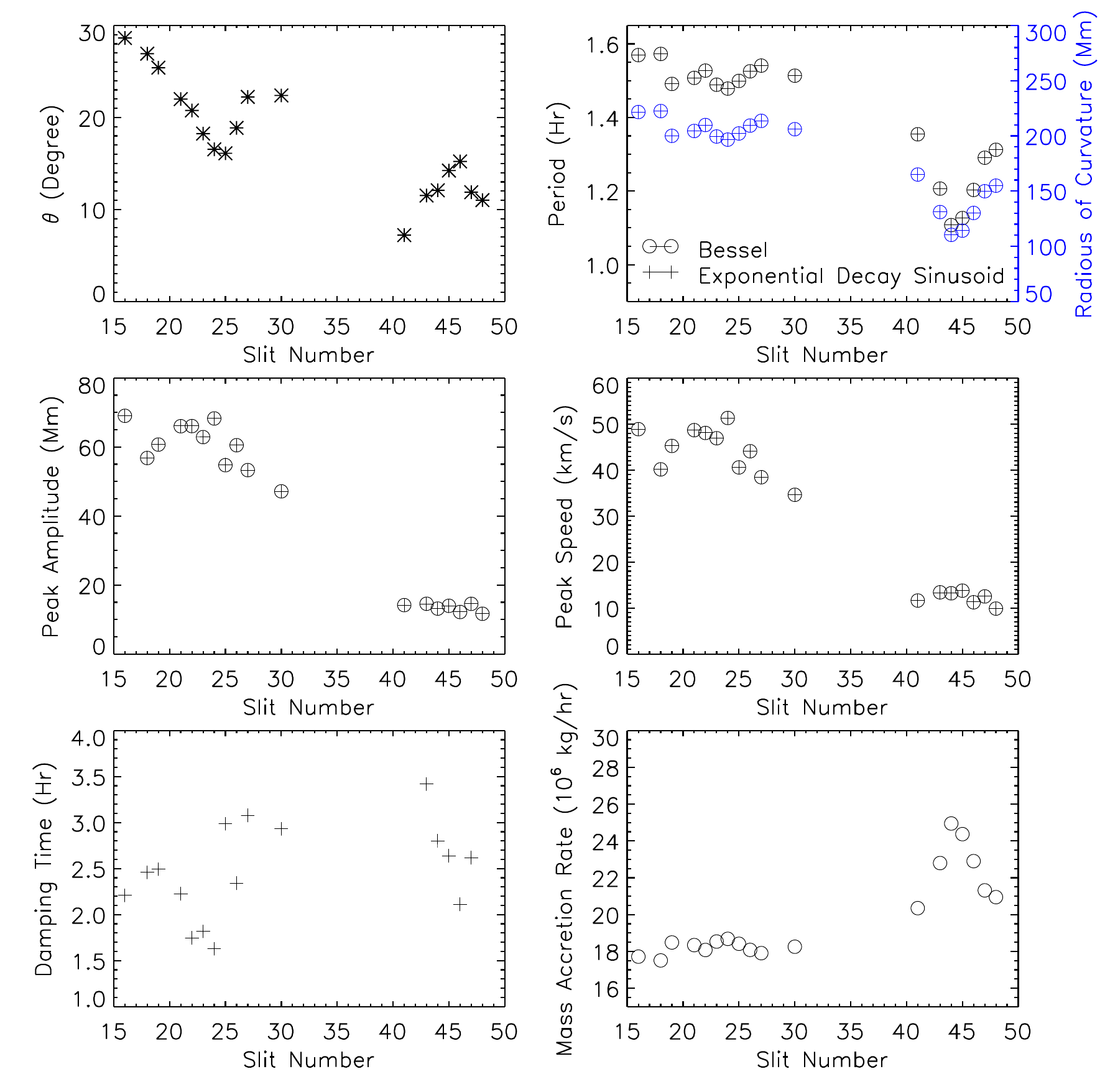}
\caption{Magnetic field parameters along the filament spine. The abscissa indicates the slit number (from east to west) centering through the 50 spine points in Figure~\ref{fig:TD_maps_along_spine}a. Panel (a) denotes the angle of oscillation direction with respect to the spine. Panels (b--f) plot the parameters obtained by fitting the observed oscillatory patterns (see Figure~\ref{fig:TD_maps_along_spine}) with the damped sinusoidal function ('+' symbol) and Bessel function (circle). In Panel (b) the curvature radius $R$ (blue) is given by $2\pi/P=\sqrt{g_\odot/R}$ (see the text) and scaled by the right $y$-axis.}
\label{fig:filament_mag_params}
\end{figure}

We then visually identified the best TD map characterizing the oscillations at each spine position, following the same rules as in \citet{Luna2014}. For each selected map, the angle of the slit relative to the spine provides the oscillation direction ($\theta$), which is found to vary in the range 7$^{\circ}$--30$^{\circ}$ (Figure~\ref{fig:filament_mag_params}b). At the intermediate spine positions, the oscillation direction ($\sim$23$^{\circ}$) is consistent with previous observations utilizing the Hanel effect to measure the field direction relative to the filament spine \citep{Leroy1983}. Next, the oscillatory pattern $y(t)$, extracted using the method elaborated in Appendix~\ref{appendix:TD_spine}, are fitted with both the exponentially decaying sinusoidal and Bessel functions as follows \citep{Luna+Karpen2012, Luna2014},

\begin{equation}\label{eq:damped_sinusoidal}
 y_{\sin}(t)=y_0+b(t-t_0)+A \sin(\frac{2\pi}{P}(t-t_0)+\phi_0)\exp(-(t-t_0)/\tau),
\end{equation}
\begin{equation}\label{eq:bessel}
 y_{\mathrm{Bessel}}(t)=y_0+b(t-t_0)+A J_0(\frac{2\pi}{P}(t-t_0)+\psi_0)\exp(-(t-t_0)/\tau_\omega),
\end{equation}
where $y_0$ is the oscillation center at the time ($t_0$) of maximum amplitude $A$. $P$, $\phi_0$, and $\tau$ represent the period, initial phase, and damping time-scale of the oscillation, respectively. $\tau_\omega$ in Eq.~\ref{eq:bessel} accounts for additional small-scale energy losses. In general, both functions yield almost the same fitting to the oscillation patterns (Figure~\ref{fig:aia_nvst_oscillation}c \& d) with a typical standard deviation $\sigma^2$ $\leq$ 50 arcsec$^2$, indicating the fitting results to be reliable. The difference between the two fitting curves is extremely small compared to the dynamic range of the oscillation (see Appendix~\ref{appendix:TD_spine}).

From the fitting results, one can see that the oscillatory pattern in the eastern section is distinct from that in the western section in terms of the oscillation direction $\theta$, period $P$, peak amplitude, and peak speed (Figure~\ref{fig:filament_mag_params}). Particularly, a simple pendulum analogy gives the curvature radius $R$ of field lines (Figure~\ref{fig:filament_mag_params}b), i.e., $2\pi/P=\sqrt{g_\odot/R}$ \citep{Luna2014}. The calculation indicates that the curvature radius of field lines in the east (200--220 [$\pm$15]~Mm\footnote{The uncertainties of all the fitting parameters are relatively small and hence not shown in Figure~\ref{fig:filament_mag_params}, but the maximum is reported alongside the respective parameter.}) is larger than that in the west (110--160 [$\pm$10]~Mm; Figure~\ref{fig:filament_mag_params}b). In the TD maps one can discern in the east an oscillatory pattern (Figure~\ref{fig:aia_nvst_oscillation}c) similar to that in the west (Figure~\ref{fig:aia_nvst_oscillation}d). It is very faint, probably obscured by the dominant oscillation in the east, but appears in phase with its western counterpart. Thus, two distinct and out-of-phase oscillatory patterns coexist in the east, but only one exists in the west.

The damping of oscillations can be modeled by a continued mass accretion by condensation at a rate $\alpha=2\pi m_0/P\psi_0$ \citep{Luna+Karpen2012}, where $m_0$ is the mass of the filament at $t_0$. $P$ and $\psi_0$ are defined in Eq.~\ref{eq:bessel}. Here the surge feeds mass into the filament. We estimated $m_0$ to be $5\pm3.8\times10^8$ kg, following Eq.~(8) in \citet{Luna2014}. We found that $\alpha$ ranges between 17--26$\times10^{6}$ kg/hr, comparable to $36\pm27\times 10^6$ kg/hr in \citet{Luna2014}. The error is large ($20\times10^{6}$ kg/hr in our case) because of large uncertainties in estimating $m_0$.

\subsubsection{Rotational Motion Around the Spine}
\begin{figure}
\centering
\includegraphics[width=0.9\textwidth, angle=0]{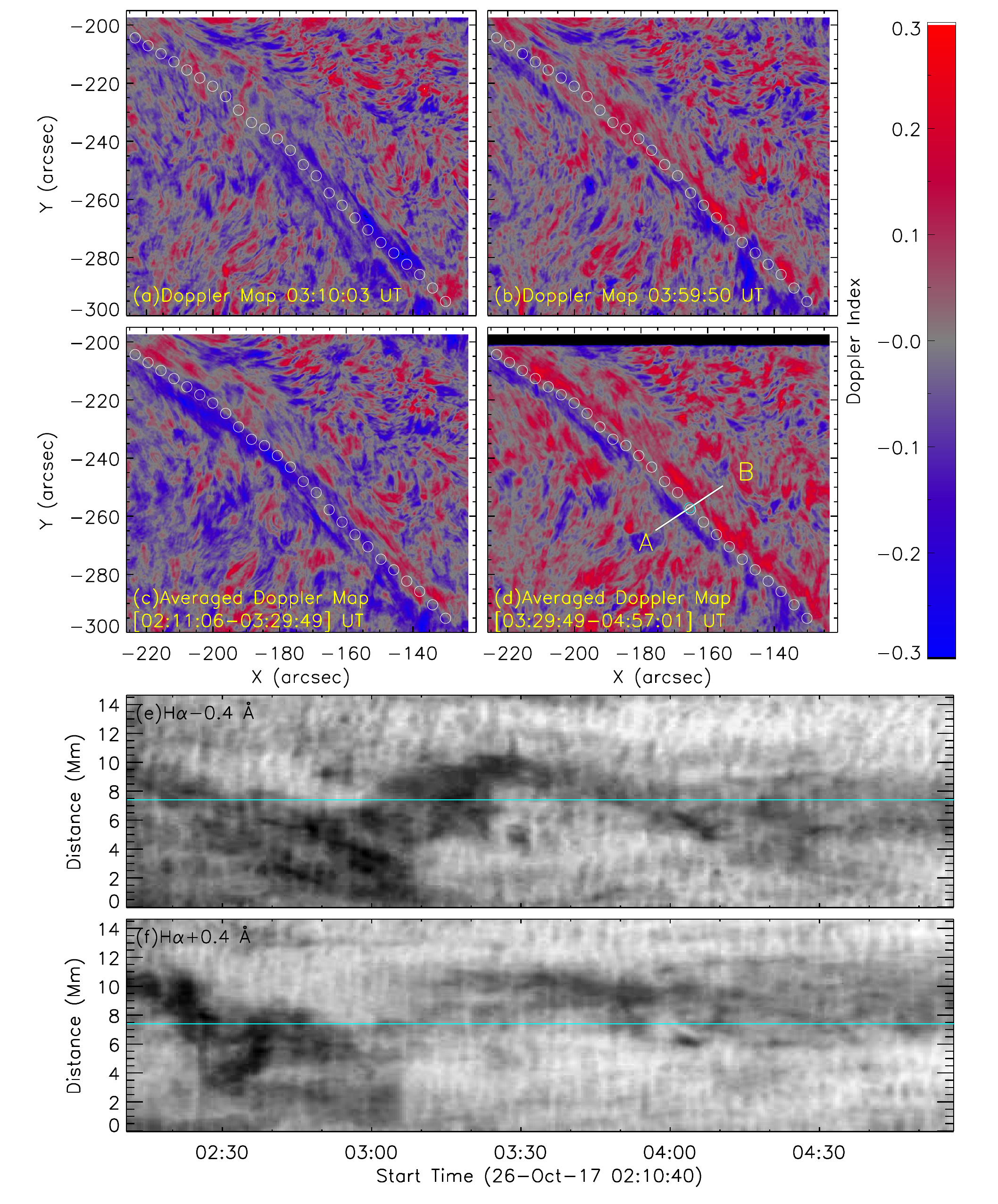}
\caption{Rotational motion around the filament spine. Panels (a--b) show Doppler maps indicating dominant upflow at 03:10:03 UT and downflow at 03:59:50 UT at the opposite sides of the spine. Panels (c--d) show Doppler maps averaged over 02:10:53--03:30:00 UT and 03:30:00 - 05:00:00 UT, respectively, aggregately confirming the rotational motion concentrated about the spine. In Panels (e--f), TD maps show an alternative appearance of material in the blue (e) and red wing (f) along the slit (A to B; marked in Panel (d)). The location of the spine is marked by circles in Panels (a)--(d) and by cyan horizontal lines in Panels (e) and (f). An animation is available online\textsuperscript{b}, with each frame consisting of H$\alpha$ images in the line center and line wing ($\pm0.4$~{\AA}) and the corresponding Doppler map.}
\small\textsuperscript{b} Available at \url{https://www.dropbox.com/s/tfez24no8yzz7ea/animation_doppler_rotation.avi?dl=0}
\label{fig:filament_rotation}
\end{figure}

Besides the longitudinal oscillations, an apparent rolling or transverse motion of individual filament threads is discernible around the filament spine, especially in the western section (see the animation accompanying Figure~\ref{fig:filament_rotation}), which is reminiscent of the helical motions studied by \citet{Okamoto2016}. To further understand this motion, we constructed Dopplergrams employing the following equation \citep{Langangen2008},
\begin{equation}
 D=\frac{B-R}{B+R}
\end{equation}
where $B$ and $R$ denote the pixel intensity recorded in the blue-wing (H$\alpha$-0.4~{\AA}) and red-wing (H$\alpha$+0.4~{\AA}) images, respectively. For absorption features like filaments, Doppler blue-shift (red-shift) is translated to  negative (positive) Doppler index $D$, indicating plasma motions towards (away from) the observer. At 03:10:03~UT the Doppler map shows an upward motion of the material predominately at the immediate south of the spine   (Figure~\ref{fig:filament_rotation}a) and at 03:59:50~UT a downward motion at the immediate north of the spine (Figure~\ref{fig:filament_rotation}b). Averaging Dopplergrams over 02:11:06--03:29:49 UT (Figure~\ref{fig:filament_rotation}c) and 03:29:49--04:57:01 UT (Figure~\ref{fig:filament_rotation}d), we found this pattern persists, i.e., consistent blue (red) shift appears in the immediate south (north) of the filament spine during the whole temporal sequence. This suggests a rotational motion of the filament material around the spine, because the Doppler pattern would have changed periodically and ended up canceling out each other with a long time integration if the filament material were oscillating up and down.

To make further sense of the Doppler pattern, we prepared time-distance maps with blue- and red-wings H$\alpha$ images (Figure~\ref{fig:filament_rotation}(e \& f)) along a slit perpendicularly across the spine (Figure~\ref{fig:filament_rotation}d). Through the slit longitudinal oscillations are recorded as transverse motions across the spine. But beside the transverse motions, one can also see that at 03:10 UT the blue wing looks significantly darker than the red wing in the south of the spine, which explains the dominant blue-shift therein (Figure~\ref{fig:filament_rotation}a). In contrast, at 03:59 UT the red wing looks darker than the blue wing in the north of the spine, which explains the dominant red-shift therein (Figure~\ref{fig:filament_rotation}b). Thus, either blue or red shift may dominate in an instantaneous Dopplergram because of the longitudinal oscillations superimposed upon the rotational motions.

Oppositely directed motions on two sides of the filament spine have been noticed before \citep[e.g.,][]{Engvold1985,Williams2009}, and were interpreted by either flows following the magnetic field of a helical flux rope \citep{Engvold1985} or internal motions of the flux-rope field lines due to the kink instability \citep{Williams2009}. With high-resolution EUV and H$\alpha$ observations presented above, we are able to exclude the latter possibility, as no kinking motions of the filament spine are visible \citep{Gilbert2007soph}.

\section{Discussion and Conclusion} \label{sec:disc_conc}
\begin{figure}
\centering
\includegraphics[width=0.8\textwidth, angle=0]{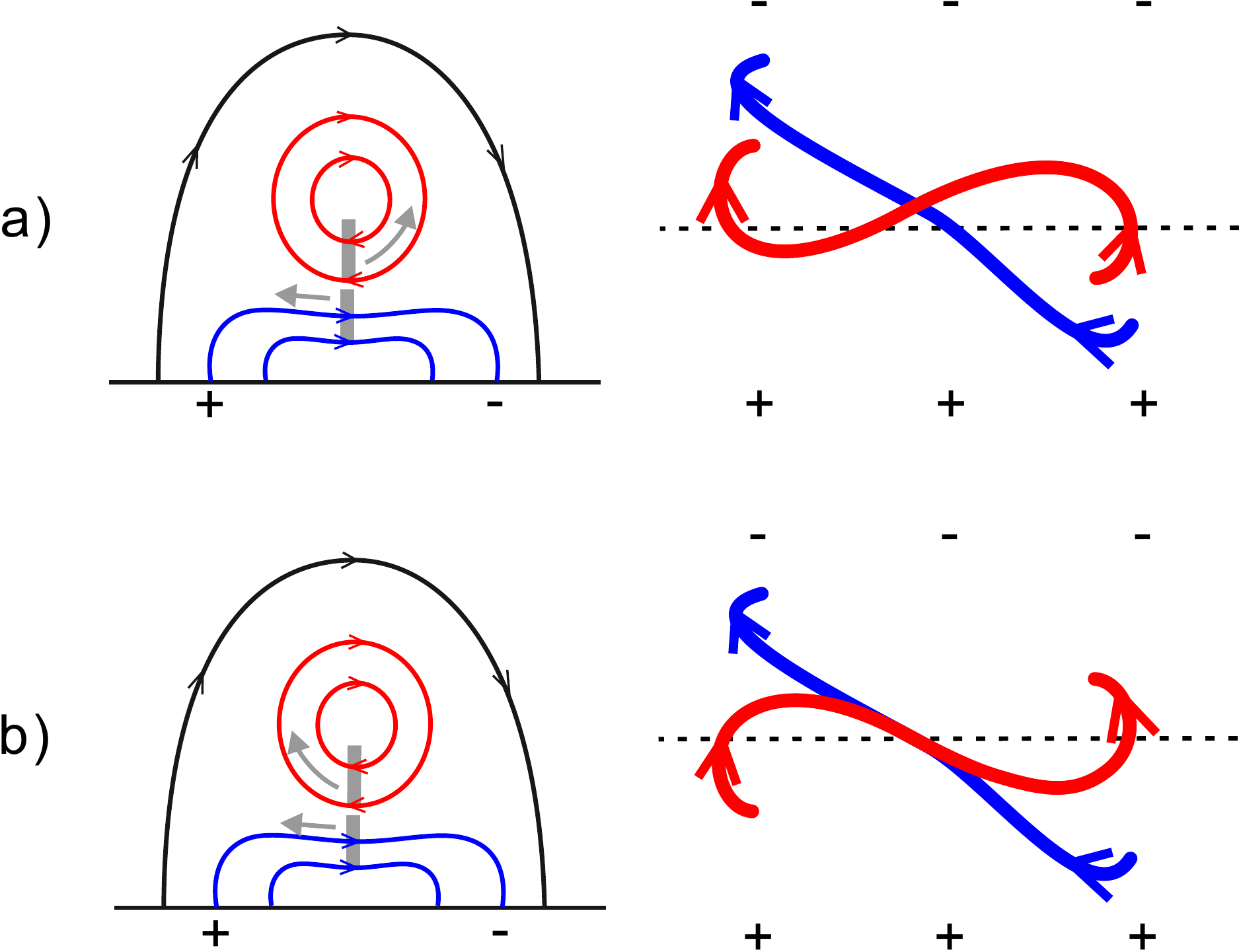}
\caption{Double-decker configuration of the magnetic field hosting the filament. In the cross section (left column) viewed from the west (solar north to the right), the configuration features a flux rope (red) atop a sheared arcade (blue). Viewed from above (right column), the flux rope and the sheared arcade possess the same (a) and opposite (b) sense of magnetic helicity, respectively. Grey arrows in the left column indicate the initial moving directions of filament mass situated at the field-line dips (denoted by grey bars), given that the filament is disturbed by a westward surge initiated in the east. The observation is consistent with the configuration in the bottom panels. }
\label{fig:filament_cartoon}
\end{figure}

This paper presents distinctive yet co-existing motions in an active-region filament, \emph{viz.,} rotation around, and longitudinal oscillation along, the filament spine. Constrained by the EUV filament channel, the large-amplitude longitudinal oscillations make an acute angle with the filament spine, which is consistent with oscillations along sheared magnetic field lines across the PIL \citep{Karpen2005, Luna+Karpen2012, Luna2018}. The rotating plasma, on the other hand, argues strongly for the presence of twisted field lines in the vicinity of the spine. Since there is no sign of bald patches \citep[Figure~\ref{fig:filament_and_flare}c;][]{Titov1993} -- an inverse-polarity configuration at the photosphere ($\mathbf{B}\cdot\nabla B_z|_\mathrm{PIL}>0$), we propose a `double-decker' configuration comprising a flux-rope atop a sheared arcade \citep[Figure \ref{fig:filament_cartoon}; see also][]{Liu2012} to account for the filament magnetic field as outlined by plasma motions. The wide EUV filament channel corresponds to the sheared arcade, while the narrow H$\alpha$ filament corresponds to the co-spatial dips of both the flux rope and of the sheared arcade, where filament material is expected to be mostly concentrated and thermally insulated. Flux ropes often have an inverted teardrop shape in cross section \citep[e.g.,][]{Jiang2018}, which may explain why the Doppler shifts associated with rotational motions are mainly detected around the spine. However, such a configuration appears not uniform along the spine. This is evidenced by the two distinctive patterns of longitudinal oscillations (Figure~\ref{fig:filament_mag_params}), from which we derive small-$R$ field lines in the western section, but both small- and large-$R$ field lines in the eastern section. Since twisted field lines are typically longer and possess smaller curvature radii than sheared field lines, we speculate that the sheared arcade in the east does not communicate with its counterpart in the west, while the flux rope extends along the whole spine as suggested by the Dopplergrams (Figure~\ref{fig:filament_rotation}). It is likely that the surge disturbs both the eastern sheared arcade and the flux rope, but fails to trigger longitudinal oscillations in the western sheared arcade.

This \emph{hybrid} configuration may accommodate two possible scenarios depending upon how the flux rope is winded and anchored. Consequently, plasma would move in opposite directions in the flux rope when disturbed by the westward surge initiated at its eastern end (Figure~\ref{fig:filament_cartoon}). Although Dopplergrams are only available during the 2nd oscillation cycle and beyond (marked in Figure~\ref{fig:aia_nvst_oscillation}(d \& e)), we hypothesize that the rotational pattern is similar during the 1st cycle, with upward (downward) motions at the immediate south (north) of the spine. This pattern is consistent with Figure~\ref{fig:filament_cartoon}b, in which the flux rope and sheared arcade are anchored at the opposite side of the PIL but the twisted field lines cross the PIL in an inverse-polarity manner. We hence exclude the alternative scenario (Figure~\ref{fig:filament_cartoon}a), in which the flux rope also has an inverse-polarity configuration but is anchored at the same side of the PIL as the sheared arcade.
	
In the proposed configuration (Figure~\ref{fig:filament_cartoon}b), the right-skewed sheared arcade possesses positive helicity, which is dominant in the southern hemisphere \citep{Pevtsov1995}; but the flux rope possesses negative (left-handed) helicity, exhibiting an inverse-S shape, the dominant shape of soft X-ray sigmoids in the northern hemisphere \citep{Pevtsov2001}. Following the reasoning by \citet{Chen2014}, one would expect right-bearing barbs at the dips of both the shear arcade and the flux rope, because the arcade and the rope possess opposite signs of helicity. Indeed the dynamic barb as excited by the surge is right-bearing; but the filament is sinistral because the axial field points to the left if we view the filament from the positive-polarity side \citep{Martin1998}. Here the correspondence between the filament chirality and the bearing sense of barbs does not conform to the conventional wisdom because mixed signs of helicity is distributed above the same PIL. Such a distribution of helicity may result from an injection of opposite helicity through photospheric evolution \citep[e.g.,][]{Liu2010,Chandra2010,Romano2011}. Unfortunately we cannot explore this possibility here because AR 12685 was highly decayed at the time of the event investigated. If there were any significant helicity injection, it must have occurred during the early phase of the active region.

Alternatively, an injection of excess twist into the filament from one footpoint could drive a combination of longitudinal and rotational plasma motions \citep{Uchida2001,Vrvsnak2007}. However, when the Alfv\'{e}n-wave packet caused by the injection is reflected at the opposite footpoint, the direction of rotational motions is expected to reverse, but in our case the Doppler pattern persists (Figure~\ref{fig:filament_rotation}). Moreover, we found no sign of twisting motions in the surge. Hence we dismissed the wave-packet interpretation for the current observation. But we cannot dismiss the possibility that the observed rotational motion is a manifestation of unwinding relaxation; in other words, magnetic twist is propagating out of rather than into the filament, which could happen when the twisted filament field reconnects with the ambient, untwisted field \citep{Okamoto2016}. Additionally, \citet{Jing2003,Jing2006} speculated that a mass perturbation in a fine thread would propagate along the spine across the obliquely stacked threads making up the filament. In our high-resolution observations, it is clear that mass motions involved in longitudinal oscillations are directed along individual threads, but it is obscure whether there are any perturbations propagating across the threads.

To conclude, this study has demonstrated a promising methodology to diagnose the magnetic configuration of filaments and has substantiated the presence of the double-decker configuration with a flux rope atop a sheared arcade \citep{Liu2012}. Since no discernible separation is seen inside the filament studied, such a configuration might be more common than previously thought. With this configuration, however, one expects normal polarity in the lower atmosphere but inverse polarity in the higher atmosphere; because of the mixed signs of helicity, one also expects two pairs of plasma draining sites if the filament fully erupts, except that double-decker filaments often erupt partially \citep[e.g.,][]{Liu2012,Cheng2014,Zhu2015}. These predictions remain to be verified by future observation.

\acknowledgments

A.K.A. acknowledges the support from the Chinese Academy of Science (CAS) as well as the International Postdoctoral Program of University of Science and Technology of China. R.L. acknowledges the support from NSFC 41474151, 41774150, and 41761134088. This investigation made use of the data acquired from NVST which is operated by Yunnan Astronomical Observatory, China. A.K.A. acknowledges the hospitality offered by the staff of Fuxian Solar Observatory during his stay for the observing period and Dr. Y. Y. Xiang for helping with the alignment of NVST images. The authors also acknowledge the free data usage policy of \textit{SDO}, \textit{GOES}, and \emph{GONG} H$\alpha$ network.

\software{MPFIT \citep{mpfit}	}

\appendix
\section{Extraction of Oscillatory Patterns} \label{appendix:TD_spine}
\begin{figure}
	\centering
    \includegraphics[width=0.99\textwidth, angle=0]{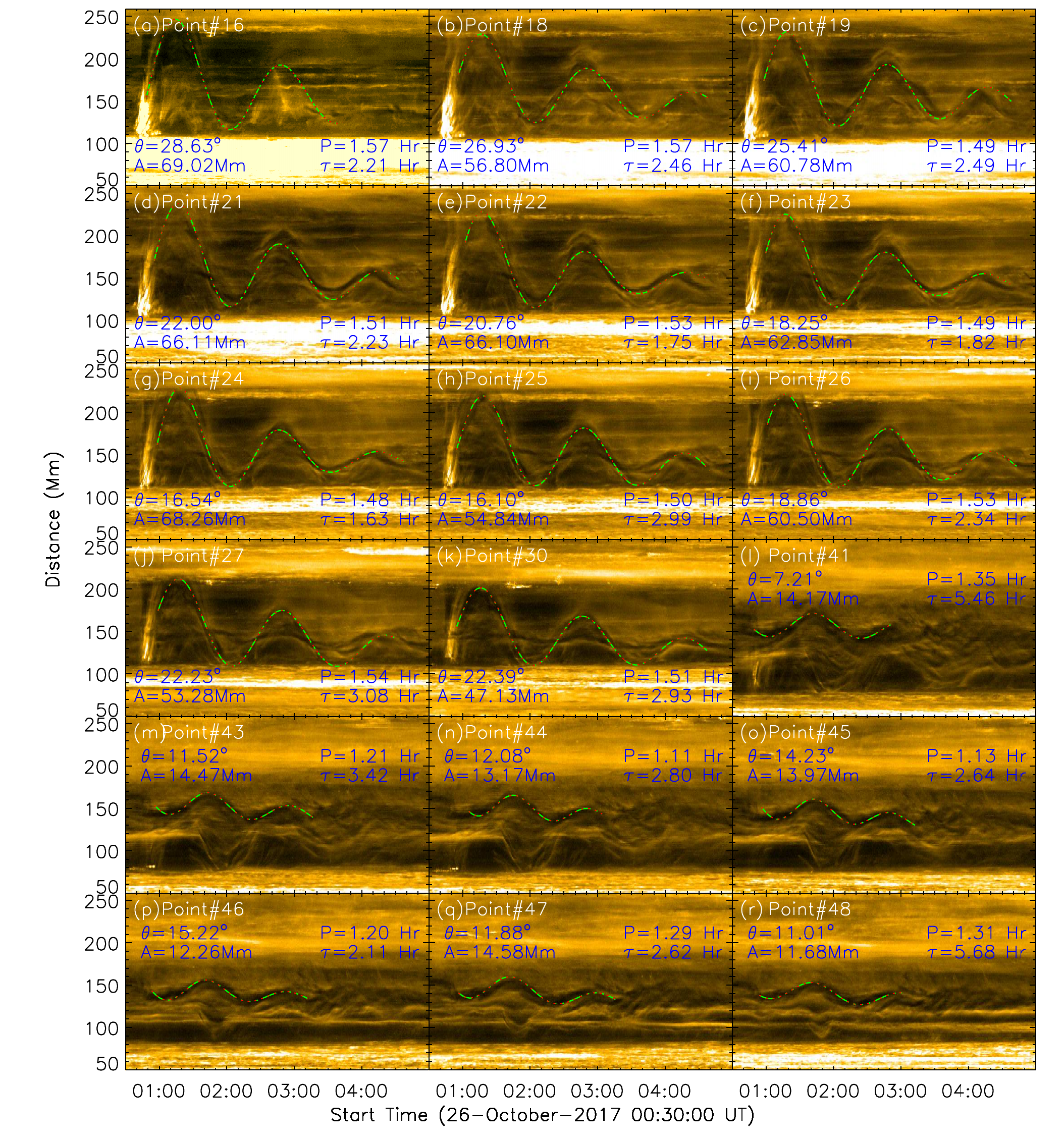}
	\caption{Selected time-distance maps characterizing the oscillations at various positions on the filament spine. TD maps taken from the eastern section of the filament spine are shown in Panels (a)--(k), while those form the western section in Panels (l--r). Overplotted are best-fit exponentially decaying sinusoidal (red) and Bessel (green) functions. The fitting parameters such as oscillation angle ($\theta$), period ($P$), amplitude ($A$), and damping time-scale ($\tau$) are annotated in the respective panels. Spine positions are marked in white.}
	\label{fig:TD_maps_along_spine}
\end{figure}

\begin{figure}
 \centering
\includegraphics[width=0.99\textwidth, angle=0]{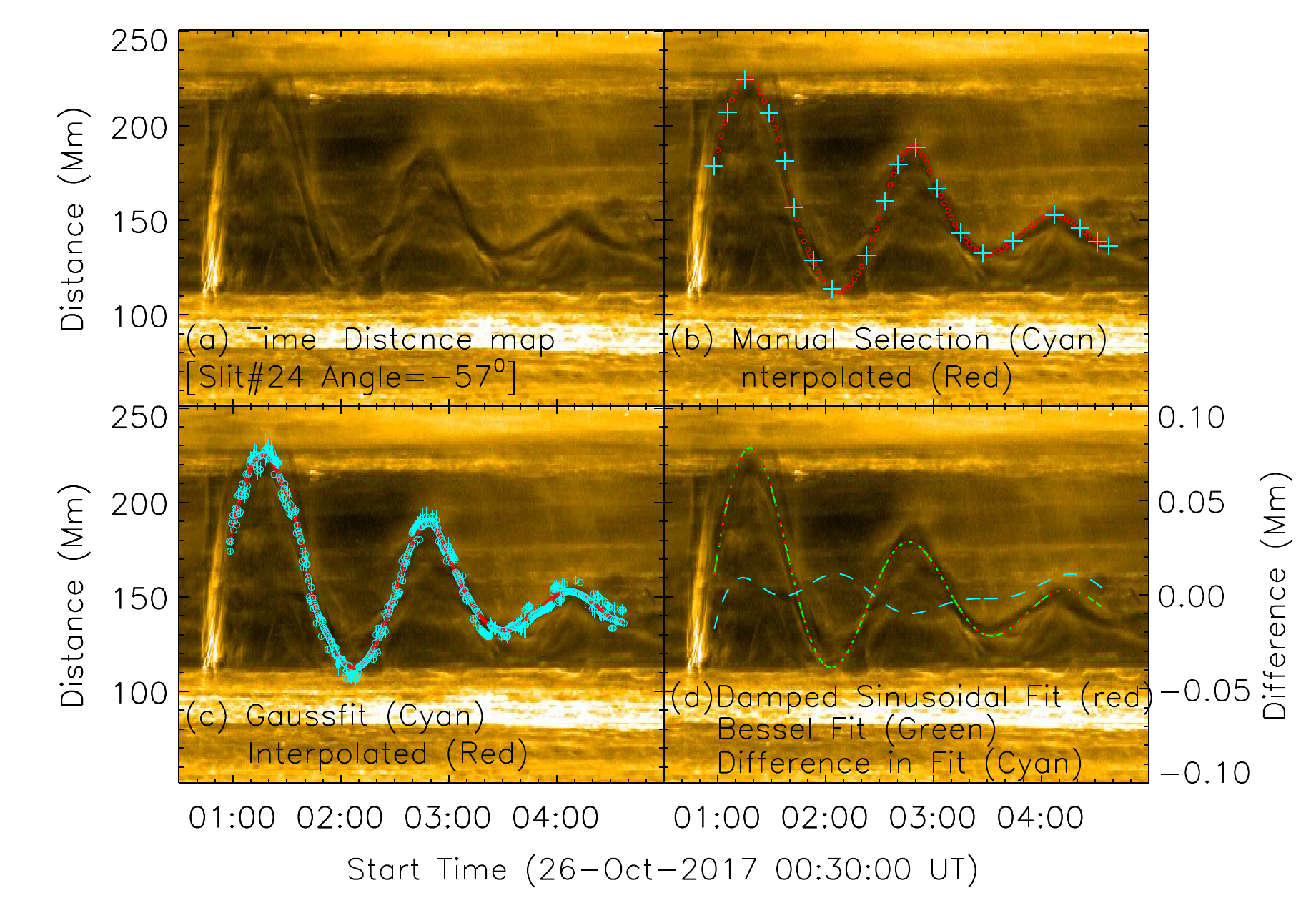}
\caption{Application of the semi-automatic technique to extract the oscillation pattern. (a) A TD map constructed at position number 24 on the filament spine, through a slit inclined at -57$^\circ$ with respect to the solar west. (b) The visible oscillation pattern is identified by manually selecting 20 points (cyan `+' symbols), based on which a cubic spline interpolation give points at all time instants (red circles). (c) The oscillation pattern is refined by applying the Gaussian fitting to locate the minimum intensity positions (cyan; see the text). (d) Overplotted are the best-fit Bessel function (green) and exponentially decaying sinusoidal function (red). Their difference is plotted in cyan and scaled by the right $y$-axis.}
\label{fig:oscl_pattern_extract_technique}
\end{figure}

From the representative TD maps selected along the spine (Figure~\ref{fig:TD_maps_along_spine}), we extracted the oscillatory pattern through a semi-automatic procedure which starts by visually choosing 20 points outlining the entire discernible pattern (Figure~\ref{fig:oscl_pattern_extract_technique}b). Next, we applied a cubic spline to interpolate the intermediate points (Figure~\ref{fig:oscl_pattern_extract_technique}b). We further refined the interpolations if the intensity profiles deduced from the TD maps can be well fitted by a Gaussian function (\texttt{gaussfit.pro}; Figure~\ref{fig:oscl_pattern_extract_technique}c). Each intensity profile is 20 pixels wide, centering on the minimum intensity location within $\pm$15 pixels of the original interpolated location. The full-width-half-maximum of the best-fit Gaussian function is considered as the uncertainty of oscillation amplitude. We then fitted the extracted oscillation pattern by the exponentially decaying sinusoidal and Bessel functions with a non-linear least squares fitting method (\texttt{mpfitfun.pro}; Figure~\ref{fig:oscl_pattern_extract_technique}d). The difference between the two fitting curves is very small with a maximum value $\pm0.02$~Mm (1 pixel on TD maps $\sim\,$0.44~Mm) for the case shown in Figure~\ref{fig:oscl_pattern_extract_technique}d.

\pagebreak

\end{document}